\documentclass[12pt,a4paper]{article}
\pdfoutput=1
\usepackage{jstyle}
\usepackage{amsmath, amsfonts, amssymb}
\usepackage{tensor}
\usepackage{graphicx, subfig, hyperref, color, verbatim}
\usepackage[dvipsnames]{xcolor}
\usepackage{float}
\usepackage{shuffle}

\newcommand{\bea}{\begin{equation}\begin{aligned}}
\newcommand{\eea}[1]{\label{#1}\end{aligned}\end{equation}}
\newcommand{\beq}{\begin{equation}}
\newcommand{\eeq}{\end{equation}}

\newcommand{\dd}{\mathrm{d}}

\title{\boldmath Tensor global symmetries and the Stueckelberg mechanism for tensor fields}

\author[a,b]{Athanasios Chatzistavrakidis,}
\author[a,c]{Arash Ranjbar}
\author[a]{and Sara Zeko}
\affiliation[a]{Division of Theoretical Physics, Rudjer Bo\v skovi\'c Institute \\ Bijeni\v cka 54, 10000 Zagreb, Croatia.}
\affiliation[b]{Institute for Theoretical Physics, University of Wrocław \\
pl. Maxa Borna 9, 50-204 Wrocław, Poland.}
\affiliation[c]{University of Rijeka, Faculty of Physics\\ Radmile Matej\v ci\'c 2, 51000 Rijeka, Croatia.}

\emailAdd{athanasios.chatzistavrakidis@irb.hr}
\emailAdd{aranjbar@phy.uniri.hr}
\emailAdd{sara.zeko@irb.hr}

\abstract{We investigate the concept of \emph{tensor global symmetries}, featuring conserved currents of mixed symmetry and higher spin Nambu-Goldstone bosons. We develop a Stueckelberg mechanism for mixed symmetry tensor fields at the linearized level, focusing on the massive graviton, the massive $(2,1)$ Curtright field and the massive $(2,2)$ field. Counting degrees of freedom, we identify the set of fields that necessarily appear in the gauge invariant Stueckelberg action in each case. These fields transform under shift symmetries and they are a vector and a scalar in the first case, a graviton, a Kalb-Ramond field and a vector in the second case and a Curtright field and a graviton in the third case. The analysis results in gauge invariant and fully conserved currents of mixed symmetry for the corresponding gauge theories, which are linked to their tensor global symmetries and they can be minimally coupled to suitable background fields. Viewing the graviton and the Kalb-Ramond field as Nambu-Goldstone bosons for constant symmetric and antisymmetric shift symmetries, we use a nonminimal coupling to uncover a 't Hooft anomaly in linearized gravity.}

\begin{document}
\maketitle
\tableofcontents


\section{Introduction}

Higher-form gauge symmetries frequently appear in theories such as supergravity and string theory, and they are long known and extensively studied \cite{10.1007/BFb0104621,Kalb:1974yc,Freedman:1976xh,Cremmer:1978km}, also from the mathematical standpoint of higher gauge theory \cite{Baez:2005qu,Baez:2010ya}. Recently,  these symmetries were revisited as global symmetries through the lens of Noether’s theorem \cite{Gaiotto:2014kfa}. In this framework, ``generalized" global symmetries extend standard symmetries by replacing rigid scalar transformation parameters with higher-form parameters. 

A direct class of examples in this direction are quantum field theories featuring vector global symmetries \cite{Hsin:2018vcg,Seiberg:2019vrp,Seiberg:2020bhn}, continuous or discrete, some of which were inspired by the related development of fracton physics \cite{Chamon:2004lew,Pretko:2020cko}.  Higher-form global symmetries exhibit interesting spontaneous symmetry breaking patterns \cite{Kapustin:2013uxa, Lake:2018dqm}. Besides higher-form symmetries, there exist many classes of quantum field theories that exhibit more intricate structures such as higher-group global symmetries and non-invertible symmetries \footnote{The rapidly growing literature on the subject and the developments prior and subsequent to \cite{Gaiotto:2014kfa} are described in several recent reviews, such as \cite{Cordova:2018cvg,Schafer-Nameki:2023jdn,Shao:2023gho,Sharpe:2015mja,Cordova:2022ruw,Benini:2018reh}.}. The gauging of these generalized global symmetries has proven to be very intriguing. This leads to a broader view: while objects charged under standard symmetries are point particles, those charged under generalized symmetries are extended objects \cite{Gaiotto:2014kfa}. Specifically, in a $p$-form symmetry, the charged object is a $p$-form field, and the invariant charge operator is defined on a co-dimension $p+1$ surface \cite{Lake:2018dqm}.

Higher-form global symmetries have been applied in various areas, including the formulation of relativistic hydrodynamics \cite{Grozdanov:2016tdf}, studies in holography \cite{Hofman:2017vwr}, and the construction of new models for topological phases of matter \cite{Yoshida:2015cia}. The latter application directly relates to the breaking of higher-form global discrete symmetries. This raises the broader question of spontaneous symmetry breaking in higher-form symmetries \cite{Hofman:2018lfz, Hirono:2022dci}. Just like spontaneous breaking of standard symmetry leads to a scalar field becoming the Goldstone boson of the broken symmetry, spontaneous breaking of $p$-form shift symmetries results in $p$-form gauge fields becoming the Goldstone bosons of the corresponding broken symmetries \cite{Lake:2018dqm}. 

This interesting outcome can also be understood through the mixed 't Hooft anomaly present in theories with higher-form symmetries. This anomaly suggests that although both higher-form electric and magnetic currents—corresponding to electric and magnetic global shift symmetries—are present, it is impossible to gauge both symmetries simultaneously. Similarly to the standard case, where the presence of a mixed 't Hooft anomaly indicates the realization of scalars as Goldstone bosons for spontaneously broken $0$-form $U(1)$ symmetry, the 't Hooft anomaly in $p$-form symmetries points to the possibility of $p$-form gauge fields becoming Goldstone bosons for broken $p$-form $U(1)$ symmetries. It is important to note that, unlike $0$-form symmetries which can be non-Abelian, higher-form symmetries are always Abelian \cite{Kapustin:2013uxa}.

A natural extension of this discussion is the generalization of $p$-form symmetries to mixed tensor symmetries (sometimes referred to as $(p,q)$-form symmetries). A notable example of a mixed symmetry tensor field is the graviton, which can be described as a $(1,1)$-symmetric field. This leads to an intriguing question: can the graviton be realized as a Goldstone boson for a spontaneously broken shift symmetry with a symmetric parameter? Recent works that explore this question are \cite{Hinterbichler:2022agn,Benedetti:2021lxj,Hull:2024bcl}, see also \cite{Cheung:2024ypq} for a 1-form symmetry in nonlinear gravity. There is an entirely different approach \cite{Borisov:1974bn, Ivanov:1976zq, Ivanov:2016lha} to describe graviton as a Goldstone boson based on the nonlinear realizations of the affine and conformal symmetries which are then spontaneously broken to the Poincare group. In \cite{Hinterbichler:2022agn}, an explanation is offered by considering the linearized theory of gravity, focusing on a subclass of general shift symmetries where the shift symmetry parameter is written as the divergence of a $(2,1)$ parameter. Here, the invariant charges of the theory are constructed from the Riemann curvature, which involves second derivatives of the fields—a departure from typical currents in QFT that involve first derivatives.

In \cite{Hull:2024bcl}, an alternative formulation of linearized gravity is proposed to accommodate a general shift symmetry for the spin-2 field. In this formulation, the electric current is conserved only on-shell, however, the dual magnetic current is unusual: it is a reducible current that is not conserved under the right derivative. Moreover, neither the electric nor magnetic currents are gauge-invariant under linearized diffeomorphism, complicating the standard gauging process, which typically involves coupling currents to background fields. However, this lack of gauge invariance is reminiscent of similar issues in the parent action formulation of linearized general relativity, where gauge invariance is restored with the help of an antisymmetric spin connection of Lorentz symmetry \cite{Boulanger:2003vs}. This suggests that incorporating a $2$-form field alongside the symmetric spin-2 field is essential to restore gauge invariance. This idea can be fully realized by using a first-order formulation of linearized gravity with a reducible $(1,1)$-tensor field, as originally shown in \cite{West:2001as} and further studied in \cite{Boulanger:2003vs}, offering an off-shell description of the electric-magnetic duality of linearized gravity. More recently, this was also extended to incorporate the duality relations of \cite{Hull:2001iu}, which include a $\theta$-term,  off-shell \cite{Chatzistavrakidis:2020kpx} and to provide a practical framework for tracking anomalies under duality  \cite{Chatzistavrakidis:2021dqg}. 

To address the unusual behavior of currents in \cite{Hull:2024bcl}, we propose a formulation inspired by the relationship between the Stueckelberg mechanism and Goldstone bosons, where we can construct gauge-invariant and conserved Noether currents. By introducing a mass term to the action of free massless gauge fields, thereby making the field massive, the action loses its gauge invariance. The Stueckelberg mechanism was introduced to restore gauge symmetry to massive actions by incorporating additional ``Stueckelberg" fields with their own shift symmetries. A classic example is the massive vector field theory, where $U(1)$ gauge invariance can be restored by adding a real scalar field that undergoes infinitesimal shifts under the gauge transformation. For scalar fields, this can be equivalently described using a complex scalar field formed from the doublet of real scalars. By assigning a real non-zero vacuum expectation value to this complex scalar field, the symmetry breaks spontaneously, and the Stueckelberg scalar becomes the Goldstone boson of the broken $U(1)$ symmetry. When the mass approaches zero, the fields decouple, and the action for the Stueckelberg field represents the action for Goldstone modes. 

This paper explores the connection between Stueckelberg fields and Goldstone bosons, demonstrating how it can be generalized to $(p,q)$ mixed symmetry tensor field theories. While $p$-form theories typically involve a single $p$-form Stueckelberg field that becomes the Goldstone boson for spontaneously broken $p$-form shift symmetry, more than one Stueckelberg field is necessary in $(p,q)$-form theories to restore gauge invariance. Each of these Stueckelberg fields can become the Goldstone boson. It depends on which generators are the ones of the broken part of the symmetry, which are in turn responsible for the non-zero vacuum expectation value of the dynamical fields.

We illustrate this in three cases, primarily through the massive Curtright theory \cite{Curtright:1980yk}, which describes a massive $(2,1)$ mixed tensor field (the other two cases start with a massive graviton and a massive $(2,2)$ field, respectively); see \cite{Zinoviev:2002ye, Zinoviev:2008ve} for earlier works on the massive mixed tensor fields and \cite{Curtright:2019wxg, OGIEVETSKY1965167} for the discussion of the massive dual spin-2 field. In this context, gauge invariance is restored by introducing a symmetric spin-2 field, an antisymmetric Kalb-Ramond field, and a vector field. The massless Curtright field possesses gauge symmetries with both symmetric and antisymmetric parameters. Spontaneous breaking of the symmetry generated by the symmetric parameter results in the realization of the graviton as the Goldstone boson of the broken symmetry. This perspective is advantageous because the $(2,1)$ Noether currents of the global shift symmetries are not only conserved on-shell but also gauge-invariant. This allows us to build a gauge-invariant theory from the ground up by coupling a free theory of Stueckelberg fields to a background Curtright field via the gauge-invariant current. Including nonminimal couplings, we identify a conserved and gauge invariant magnetic current, whose coupling generates a 't Hooft anomaly for the symmetric shift symmetry. 

The organization of the paper is as follows. Section \ref{section dof} is a preparatory one, where we perform a degree of freedom count that leads to the identification of the three interesting cases to be studied. In Section \ref{section forms}, we revisit the case of free Abelian gauge theory for $p$-forms, recalling the conserved currents for its two 1-form global symmetries, their coupling to background fields, and its 't Hooft anomaly. In Section \ref{tensor GS 1}, we start from the Fierz-Pauli theory for a massive graviton and apply the Stueckelberg mechanism to identify the Stueckelberg fields (a photon and a scalar). This leads to an unconventional form of Abelian gauge theory, where a gauge invariant, \emph{symmetric} conserved current is identified, which can be coupled to a symmetric 2-tensor background field. This is the first case of a higher tensor global symmetry. In Section \ref{section tensor GS 2}, we start with a massive Curtright field and determine its Stueckelberg action, in terms of three Stueckelberg fields, a graviton, a Kalb-Ramond field and a vector. We identify the gauge invariant, fully conserved currents that correspond to a tensor global symmetry of type $(2,1)$. In this case, the graviton together with the Kalb-Ramond field are interpreted as Nambu-Goldstone bosons for a spontaneously broken tensor global symmetry that corresponds to symmetric and antisymmetric shifts for the two fields. Section \ref{section exotic} contains an analysis of an exotic Stueckelberg mechanism, where the massive field is a $(2,2)$ mixed symmetry tensor field and the Stueckelberg fields are a Curtright field and a graviton. In Section \ref{section anomalies}, we return to the case of the graviton as a Nambu-Goldstone boson and investigate the presence of a 't Hooft anomaly, identifying the magnetic current which is responsible for this feature. Finally, Section \ref{section conclusions} contains our conclusions and outlook.

\paragraph*{Notations.} We denote scalar fields as $\phi$ and scalar (rigid or gauge) transformation parameters as $\varphi$. 
Low-degree differential forms are denoted by their order in the Latin alphabet according to their degree; $a, b, c$ for $1$-, $2$-, $3$-forms,  reserved for Stueckelberg fields, and $A, B, C$ for $1$-, $2$-, $3$-forms, reserved for background gauge fields or massive tensor fields. Transformation parameters follow the same pattern in the Greek alphabet, $\alpha, \beta, \gamma$ for $1$-, $2$-, $3$-form parameters. For mixed symmetry tensor fields of degree $(p,q)$, $p, q\ge 0$ and $p\ge q$, we use the general notation $\omega_{p,q}$ and $\Omega_{p,q}$ for massless and massive fields, respectively, apart from the special cases of graviton ($p=q=1$) and Curtright field ($p=2, q=1$), when we use $h$ and $t$ when they are Stueckelberg fields, $H$ and $T$ when they are background or massive fields. The number of degrees of freedom of a massive tensor field $\omega_{p,q}$ is $|\omega_{p,q}|$ and accordingly $|\omega_{p,q}|_{0}$ for a massless one. We will also write $|\omega|+|\omega'|:=|\omega+\omega'|$ etc.

\section{Count of degrees of freedom}
\label{section dof}

In Proca's formulation of massive 1-forms the Lorentz gauge condition follows directly from the equations of motion. This is true strictly for nonvanishing mass and reflects the fact that the massive 1-form has three polarizations in 4D, in contrast to the two polarizations of the massless one. The Proca Lagrangian, 
\begin{equation}
    L=-\frac 14 F_{\mu\nu}F^{\mu\nu} +\frac 12 m^{2}A_{\mu}A^{\mu}\,, \label{proca}
\end{equation}
is not gauge invariant since the second term does not permit transformations of the type $\delta A_{\mu}=\partial_{\mu}\varphi$. The Stueckelberg mechanism restores this gauge invariance at the cost of introducing an additional massless scalar field $\phi$.  What effectively happens is that in $D$ dimensions the total number of degrees of freedom increases to $D+1$, while the gauge condition and the equations of motion reduce it to the $D-1$ physical polarizations. The Stueckelberg scalar is an alternative way to describe Nambu-Goldstone bosons, both being fields invariant under a shift symmetry (derivatively coupled). Then the Stueckelberg Lagrangian in the unitary gauge manifests the Higgs mechanism in the decoupling limit of the Higgs boson.

Our aim here is to generalize this idea to mixed symmetry tensor fields. Such higher tensor fields will play the dual role of the massive 1-form and of the Stueckelberg scalar. For example, we are interested in thinking of a graviton (a symmetric 2-tensor field) either as a massive field, as in Fierz-Pauli theory, or as a Stueckelberg field of a tensor gauge theory for a higher massive field, e.g., a massive Curtright field. To understand how to do this properly, it is important to have a correct degree of freedom count. 
In the ordinary Stueckelberg formalism, this is very simple; a massless scalar has the same number of degrees of freedom as the difference between the degrees of freedom of a massive and a massless  1-form: $|\phi|_{0}=|A|-|A|_0 = (D-1)-(D-2)=1$.

This motivates the following question. For a given collection of $N$ massless fields, can we write the number of their on-shell degrees of freedom as the difference between the number of massive and massless degrees of freedom for another collection of $M$ fields of higher order? This is expressed as 
\begin{equation}\label{eq:matching_dof}
\sum_{i=1}^{N}|\omega_{p_i,q_i}|_0=\sum_{j=1}^{M}(|\Omega_{p_j,q_j}|-|\Omega_{p_j,q_j}|_{0})\,, 
\end{equation}
with $\text{max}(p_j+q_j) > \text{max}(p_i+q_i)$. The left-hand side will then correspond to a collection of Stueckelberg fields and the right-hand side to a collection of massive tensor fields. This question is general and we will not answer it in full generality, focusing only on the physically most interesting cases. 

To explore the possibilities, we recall the following general counting of degrees of freedom. The independent components of an irreducible tensor field that corresponds to an irreducible two-column Young tableau of order $(p,q)$ in $D$ dimensions, are given by the formula: 
\begin{equation}
    \textrm{dim}_D(p,q) = \left(\begin{array}{c}
        D  \\ p 
    \end{array}\right)\left(\begin{array}{c}
        D+1  \\ q 
    \end{array}\right)\left(1-\frac{q}{p+1}\right)\,.
\end{equation}
Taking into account gauge fixing conditions and equations of motion, the on-shell degrees of freedom of a general mixed symmetry tensor field are
\begin{align}
    & |\omega_{p,q}| = \textrm{dim}_{D-1}(p,q)- \textrm{dim}_{D-1}(p-1,q-1)\,,\\[4pt]
   & |\omega_{p,q}|_0 =   \textrm{dim}_{D-2}(p,q)- \textrm{dim}_{D-2}(p-1,q-1)\,,\label{eq:masslessonshell}
\end{align}
in the massive and massless cases respectively. For $p$-forms ($q=0$) these simplify to 
\begin{align}
    |\omega_p|=\left(\begin{array}{c}
        D-1  \\ p 
    \end{array}\right)\,, \quad |\omega_p|_{0}=  \left(\begin{array}{c}
        D-2  \\ p 
    \end{array}\right)\,.
\end{align}

Suppose that we have a single Stueckelberg field ($N=1$). When it is a $p$-form $\omega_{p}$, the unique answer to the question we posed above is 
\begin{equation}
    |\omega_p|_0=|\Omega_{p+1}|-|\Omega_{p+1}|_0\,,
\end{equation}
and the massive field is a single $(p+1)$-form $\Omega_{p+1}$ ($M=1$). For $p=0$, the usual count of the ordinary Stueckelberg mechanism is recovered and for $p=1$ it reduces to the case of the Stueckelberg mechanism for a massive 2-form, equivalently to the statement that a photon can be interpreted as a Nambu-Goldstone boson of a spontaneously broken 1-form global shift symmetry in Abelian gauge theory without sources \cite{Gaiotto:2014kfa}. For $p>1$, it results in the analogous statements for higher degree $p$-form fields. 

When the single field is not a $p$-form but a genuine $(p,q)$ tensor field, it is impossible to express it as the required difference. This suggests that there is no canonical Stueckelberg mechanism for a single $(p,q)$ tensor field. Specializing to $(1,1)$, this says that the graviton alone cannot be a Stueckelberg field for the symmetric constant shift symmetry $\delta h_{\mu\nu}=s_{\mu\nu}$; as we will argue below, this requires more fields. 

Suppose now that we start with two fields of different degrees that exhibit shift symmetries. The simplest case is a scalar field $\phi$ and a 1-form $a$. In this case, the question we posed does not have a unique answer. Specifically, 
\begin{equation}
    |\phi+a|_{0}= \begin{cases}
      \,  |A+B|-|A+B|_{0} \\[4pt] \, |H|-|H|_{0}
    \end{cases}\,.
\end{equation}
This indicates that there exist two ways to accommodate the physical degrees of freedom for the scalar/1-form system in the Stueckelberg mechanism. Either in the straightforward way of having a system of massive 1-form/2-form fields or in the more subtle way of having a massive graviton, cf. \cite{Arkani-Hamed:2002bjr}. It is one of our main goals to explain this dualism. 

If we start with three massless fields of different degrees that exhibit shift symmetries, we find for $p>q\ge 1$ the general formula  
\begin{equation}
    |(p-1,q)+(p,q-1)+(p-1,q-1)|_{0}=|(p,q)|-|(p,q)|_{0}\,,
\end{equation}
where we denoted the fields by their degree.
Read from right to left, this tells us that applying the Stueckelberg mechanism to a single massive $(p,q)$ tensor field, the massless Stueckelberg fields will be three fields. Taking as a basic example the 
case $p=2$ and $q=1$ and according to the notations explained in the Introduction, this expression reads as 
\begin{equation}
    |h+b+a|_{0}=|T|-|T|_{0}\,,
\end{equation}
namely, the graviton can be seen as a Stueckelberg field only accompanied by a Kalb-Ramond 2-form $b$ and a 1-form $a$. Another of our main goals is to explain this in detail. Our last example is the case $p=q=2$. This is an exotic tensor field with the symmetry structure of the Riemann tensor, and the above formula reads 
\begin{equation}
    |t+h|_{0}=|\Omega_{2,2}|-|\Omega_{2,2}|_{0}\,,
\end{equation}
and it tells us that the Stueckelberg fields will be a Curtright field and a graviton. 

We observe that this simple degree of freedom count assigns three distinct roles to the graviton: it can appear as (i) the gauge field coupled to a suitable current for a scalar/1-form theory, (ii) the Nambu-Goldstone boson, together with a 2-form and a 1-form, for a Curtright gauge theory, and (iii) as a ``secondary'' Stueckelberg boson for an exotic $(2,2)$ gauge theory.  

\section{Higher form shift symmetries} 
\label{section forms}

We begin with the simplest and most direct case of $p$-forms. There are two complementary ways to describe the physics. The first is to start with a massive $(p+1)$-form $\Omega:=\Omega_{p+1}$ and apply the Stueckelberg mechanism to express its Lagrangian in a gauge invariant way using a massless $p$-form field $\omega:=\omega_p$. The reverse way is to start with a massless $p$-form with shift symmetry, identify $(p+1)$-form conserved currents and couple them to $(p+1)$-form background fields. The second approach is more general and we will follow it in this section, commenting on the first approach in the end. However, we will see in the next section that the first approach is essential to get the right physics for mixed symmetry tensor fields. 

The massless $p$-form has a shift symmetry with rigid parameter $\varepsilon:=\varepsilon_{p}$, which we write directly in differential form notation, $\delta\omega=\varepsilon$. For simplicity, we will work in dimensions $D=2p+2$, where the field admits a twisted self-dual (``magnetic'') description in terms of another $p$-form $\widehat\omega$. In the electric formulation, the free action without sources is 
\begin{eqnarray}
    S[\omega]&=& -\frac 1{2}\int   \dd\omega\wedge\ast \,\dd\omega 
    =-\frac 1{2(p+1)!}\int\dd^{D}x\, F_{\mu_1\dots\mu_{p+1}}F^{\mu_1\dots\mu_{p+1}} \,,
\end{eqnarray}
where $\dd$ is the de Rham differential and the $(p+1)$-form field strength is $F=\dd\omega$. 
There are two $p$-form global symmetries (electric and magnetic) with conserved currents 
\begin{equation}
    J_{\text{e}}= \dd \omega\,, \quad J_{\text{m}}=\ast\, \dd\omega\,.
\end{equation}
The conservation of the electric current follows from the Euler-Lagrange equations of the theory (on-shell), whereas the conservation of the magnetic current is due to the Bianchi identities.

We can couple background fields $\Omega_{\text{e}}$ and $\Omega_{\text{m}}$ to the theory through these conserved currents, both $(p+1)$-forms, via the usual terms $J_{\text{e}}\wedge\ast\,\Omega_{\text{e}}$ and $J_{\text{m}}\wedge\ast\,\Omega_{\text{m}}$. To make the kinetic term invariant under the shift symmetry of the field $\omega$, we must add a ``seagull'' counterterm $\Omega_{\text{e}}\wedge\ast\, \Omega_{\text{e}}$, alluding to a mass term for the electric background field in the Stueckelberg formulation. The resulting action coupled to background fields is 
\begin{equation}
S[\omega,\Omega_{\text{e}},\Omega_{\text{m}}]= \int \bigg(-\frac 12(\dd\omega-\Omega_{\text{e}})\wedge\ast(\dd\omega-\Omega_{\text{e}})+\Omega_{\text{m}}\wedge \dd\omega\bigg)\,.
\end{equation}
The second term is a topological, BF type coupling.
The background fields have gauge transformations $\delta_{\text{e}}\Omega_{\text{e}}=\dd\varepsilon_{\text{e}}$ and $\delta_{\text{m}}\Omega_{\text{m}}=\dd\varepsilon_{\text{m}}$, where $\varepsilon_{\text{e}}$ is identified with the shift $\varepsilon$, which is now spacetime dependent. Under the magnetic transformation, the action is invariant up to a total derivative, whereas under the electric one the kinetic term is invariant but the second term gives rise to a shift: 
\begin{equation}
    \delta_{\text{e}}S=\int \varepsilon_{\text{e}}\wedge\dd\Omega_{\text{m}}\,,\label{anomaly p form}
\end{equation}
up to a total derivative. As explained in \cite{Cordova:2018cvg}, this signals a mixed 't Hooft anomaly for the product of $p$-form global symmetries, as a result of which they cannot be gauged simultaneously. The anomaly polynomial is given as 
\begin{equation}
    \mathcal I=\int \dd\Omega_{\text{m}}\wedge \dd \Omega_{\text{e}}\,,
\end{equation}
inflowing to the anomaly above via the descent equations. 

If we want to gauge the electric symmetry, we may set the background field $\Omega_{\text{m}}$ to zero, which we can do since it is not yet a dynamical field that would be integrated in the path integral. Then we should add the kinetic term for the field $\Omega_{\text{e}}$ and write the action: 
\begin{equation}
    S[\omega,\Omega_{\text{e}}] = -\frac 12 \int\bigg((\dd\omega-\Omega_{\text{e}})\wedge\ast(\dd\omega-\Omega_{\text{e}})+ \dd\Omega_{\text{e}}\wedge\ast\,\dd\Omega_{\text{e}}\bigg)\,,
    \label{p-form Stuckelberg}
\end{equation}
which is gauge invariant under the symmetry generated by the spacetime-dependent parameter $\varepsilon_{\text{e}}$. 
It is easily recognizable that \eqref{p-form Stuckelberg} is a generalized Stueckelberg action. For $p=0$ it is the usual one for a massive 1-form, and fixing the gauge by setting $\omega$ (in this case, a scalar) to zero, namely taking the unitary gauge, results in the Proca action \eqref{proca}. This could be described starting directly from the generalization of the Proca action for a massive $(p+1)$-form $\Omega$ with unit mass $m=1$ and performing the field redefinition 
\begin{equation}
    \Omega \to \Omega-\dd \omega\,.
\end{equation}
The field strength remains unchanged, as it does not see the field $\omega$, and the action \eqref{proca} is recovered. Note though that this approach does not reveal the richer structure of conserved currents and generalized global symmetries, it does not give any information on 't Hooft anomalies, and it is not suitable for electric/magnetic duality, which is not a feature of the massive theory. On the other hand, it is a guiding formulation when it is not a priori clear how gauging works, as we will see below.   

\section{Tensor shift symmetries I}
\label{tensor GS 1}

According to the degree of freedom count of Section \ref{section dof}, there exist two ways to view the system of shift symmetric scalar and 1-form $(\phi,a)$ as a pair of Stueckelberg fields. The first is obvious and does not present any new features in the analysis of the previous section. We can start either from a system of massive 1-form and 2-form $(A,B)$ and write the theory in a gauge invariant way using $\phi$ and $a$, or we can start from the canonical kinetic terms of the massless fields and gauge their electric global symmetries. However, there is an alternative path that involves a massive graviton, where the role of $(\phi, a)$ is different. This second case is what we explore in this section. 

\subsection{Stueckelberg action for the massive graviton}
\label{subsection massive graviton}

If we start with a massless scalar and a massless 1-form and ask how to couple a graviton to some conserved current associated to them, we immediately face the question of what a symmetric current is in this case. The Stueckelberg mechanism comes to the rescue; therefore, we reverse the order in our analysis, and unlike Section \ref{section forms}, we begin with the massive field. At the linearized level, a massive graviton $H_{\mu\nu}$ in 4D (for simplicity we set the mass to $1$) is described by the Fierz-Pauli action 
\begin{eqnarray}
  &&  S_{\text{FP}}=-\frac 12 \int \dd^{4}x\,\bigg(\frac 12 \partial_{\mu}H_{\nu\rho}\partial^{\mu}H^{\nu\rho}-\partial_{\mu}H_{\nu\rho}\partial^{\nu}H^{\mu\rho} +\,\partial_{\mu}H^{\mu\nu}\partial_{\nu}H^{\rho}{}_{\rho}-\frac 12 \partial_{\mu}H^{\nu}{}_{\nu}\partial^{\mu}H^{\rho}{}_{\rho}  \nonumber\\[4pt]  && \hspace{100pt} -\, \frac 12 H^{\mu\nu}H_{\mu\nu}+\frac 12 H^{\mu}{}_{\mu}H^{\nu}{}_{\nu}\bigg) \nonumber \\[4pt] && \hspace{22pt}= -\frac 14\int \dd^{4}x\int_{\text{B}}\bigg(\dd H\star \dd H-H\star H\bigg)\,. \label{fierz pauli}
\end{eqnarray}
In the last line, we expressed the action in a compact and suggestive form. It is based on introducing anticommuting coordinates that are integrated out via Berezin integration $\int_{\text{B}}$. The mathematical details are described in \cite{Chatzistavrakidis:2019len}, inspired and closely related to the bi-form formalism of \cite{Bekaert:2002dt,Bekaert:2006ix} and \cite{deMedeiros:2002qpr,deMedeiros:2003osq}, which transfers differential form notation to tensors of mixed symmetry.  Since we use this in an auxiliary way in this paper and always write component expressions at the end, we only briefly explain the basic definitions for what appears in the action \eqref{fierz pauli} and other related actions. 

The formalism uses two sets of $D$ anticommuting coordinates $\theta^{\mu}$ and $\widetilde{\theta}^{\mu}$ in $D$ spacetime dimensions. By convention\footnote{This is usually called the Deligne convention. In the present case, there is a $\mathbb{Z}_2\times\mathbb{Z}_2$ grading in which the two sets have respective degrees $(1,0)$ and $(0,1)$. The Deligne convention amounts to summing the product of each $\mathbb{Z}_{2}$ degree in the power of $-1$, which currently results in commuting sets. Anticommuting sets would be obtained with the Bernstein-Leites convention of taking instead the product of the sum of the degrees. Results do not depend on the choice of convention.} we also assume that they are mutually commuting, 
\begin{equation}
    \theta^{\mu}\theta^{\nu}=-\theta^{\nu}\theta^{\mu}\,,\quad \widetilde{\theta}^{\mu}\widetilde{\theta}^{\nu}=-\widetilde{\theta}^{\nu}\widetilde{\theta}^{\mu}\,,\quad \theta^{\mu}\widetilde{\theta}^{\nu}=\widetilde{\theta}^{\nu}\theta^{\mu}\,.
\end{equation} 
The de Rham differential for each set is defined as $\dd=\theta^{\mu}\partial_{\mu}$ and $\widetilde{\dd}=\widetilde{\theta}^{\mu}\partial_{\mu}$. For a tensor $\omega$ of type $(p,q)$ in an irreducible representation of the general linear group, we have the superfield expansion
\begin{equation}
    \omega=\frac{1}{p!q!}\omega_{[\mu_1\dots\mu_p][\nu_1\dots\nu_q]}(x)\theta^{\mu_1}\dots\theta^{\mu_p}\widetilde{\theta}^{\nu_1}\dots\widetilde{\theta}^{\nu_q}\,.
\end{equation}
The tilde can be understood as an operation on $\omega$, also, by exchange of all $\theta^{\mu}$ and $\widetilde{\theta}^{\mu}$ and resulting in a $(q,p)$ tensor $\widetilde{\omega}$. The generalized Hodge duality operator $\star$ is an operator that acting on $\omega$ results in a $(D-p,D-q)$ tensor given as 
\begin{equation}
    \star\,\omega = \frac 1{(D-p-q)!}\eta^{D-p-q}\,\widetilde{\omega}\,,
\end{equation}
where $\eta=\eta_{(\mu\nu)}\theta^{\mu}\widetilde{\theta}^{\nu}$ is the Minkowski metric thought of as a degree $(1,1)$ (super)function. This gives a good inner product that can be integrated. Berezin integration is taken for both anticommuting sets, namely 
\begin{equation}
\int_{\text{B}}:=\int\dd^{D}\theta^{\mu}\dd^{D}\widetilde{\theta}^{\mu}\,.
\end{equation}
To give an example of the action of the generalized $\star$-operator, $\star\,\dd H$ in 4D is a bi-degree $(2,3)$ tensor given as 
\begin{equation}
    \star \,\dd H= \eta_{\mu_1\nu_1}\partial_{\nu_2}H_{\mu_2\nu_3}\theta^{\mu_1}\theta^{\mu_2}\widetilde{\theta}^{\nu_1}\widetilde{\theta}^{\nu_2}\widetilde{\theta}^{\nu_3}\,.
    \label{star dH}
\end{equation}
Concatenated with the $(2,1)$ tensor $\dd H$, a maximal $(4,4)$ tensor is obtained, which is integrated with the rules of Berezin integration.

The Fierz-Pauli action is not invariant under linearized diffeomorphisms 
\begin{equation}\label{eq:trans_H}
    \delta H_{\mu\nu}=\partial_{\mu}\alpha_{\nu}+\partial_{\nu}\alpha_{\mu}\,,
\end{equation}
due to the mass term. It is also not invariant under the two-derivative transformations 
\begin{equation}\label{eq:trans_H2}
    \delta H_{\mu\nu}=-2\partial_{\mu}\partial_{\nu}\varphi\,,
\end{equation}
with a scalar parameter,
whose significance will become clear below. To restore gauge invariance, we can introduce a scalar field $\phi$ and a 1-form $a$ with the shift symmetries 
\begin{equation}\label{eq:trans_phi_a}
    \delta\phi=\varphi \quad \text{and} \quad \delta a_{\mu}=\alpha_{\mu}\,.
\end{equation}
We may now redefine 
\begin{equation}
    H_{\mu\nu} \to \mathring{F}_{\mu\nu}:=H_{\mu\nu}-\partial_{\mu}a_{\nu}-\partial_{\nu}a_{\mu}+2\partial_{\mu}\partial_{\nu}\phi\,. \label{redefined graviton}
\end{equation}
We denoted this quantity as $\mathring{F}_{\mu\nu}$ to highlight its apparent similarity to the Faraday tensor, while also emphasizing its difference, as it is not an antisymmetric tensor but a symmetric one. This tensor is gauge invariant under all symmetries with parameters $\varphi$ and $\alpha_{\mu}$: 
\begin{equation}
    \delta \mathring{F}_{\mu\nu}=0\,.
\end{equation}
In the spirit of the Stueckelberg mechanism, substituting this into the Fierz-Pauli action results in a gauge invariant action for the massive graviton at the expense of introducing the additional massless fields. Since substituting this into the expanded form of the action gives a rather complicated result, let us present it in the compact form introduced above. This only requires to write \eqref{redefined graviton} as
\begin{equation}
    \mathring{F}=H-\dd \widetilde{a}-\widetilde{\dd}a+2\dd\widetilde{\dd}\phi\,.
\end{equation}
Note that $\dd \mathring{F}=\dd H-\dd\widetilde{\dd}a$.  
The gauge invariant action takes the form 
\begin{eqnarray}
  &&  S_{\text{FP}}^{\text{St}}=-\frac 14\int\dd^{4}x\int_{\text{B}}\bigg((\dd H-\dd\widetilde{\dd}a)\star (\dd H-\dd\widetilde{\dd}a) \nonumber \\ 
    &&\hspace{110pt} -\, (H-\dd \widetilde{a}-\widetilde{\dd}a+2\dd\widetilde{\dd}\phi)\star (H-\dd \widetilde{a}-\widetilde{\dd}a+2\dd\widetilde{\dd}\phi)\bigg)\,. \qquad
\end{eqnarray}
Several terms in this action are total derivatives, specifically $\dd H\star \dd\widetilde{\dd}a, \dd\widetilde{\dd}a\star \dd \widetilde{\dd}a, \dd\widetilde{\dd}\phi\star\dd\widetilde{a}, \dd\widetilde{\dd}\phi\star \dd\widetilde{\dd}\phi$ and $\dd\widetilde{a}\star \widetilde{\dd}a$. 
Dropping these total derivatives, the action is equivalent to the Stueckelberg action
\begin{equation}
S_{\text{FP}}^{\text{St}}=\int\dd^{4}x\int_{\text{B}}\bigg(-\frac 14\big(\dd H\star \dd H-H\star H\big)+\frac 12 \big(\dd\widetilde{a}\star\dd\widetilde{a}+2H\star\dd\widetilde{\dd}\phi-2H\star\dd\widetilde{a}\big)\bigg).
\end{equation}
The first parenthesis is the Fierz-Pauli action \eqref{fierz pauli}. After these manipulations, we may perform the Berezin integration and write the final form of the remaining terms in components up to total derivatives: 
\begin{equation}
  S_{\text{FP}}^{\text{St}}-S_{\text{FP}}= \int\dd^{4}x \bigg(\frac 14 F_{\mu\nu}F^{\mu\nu}-\partial_{\mu}a_{\nu}H^{\mu\nu}+\, \partial^{\mu}a_{\mu}H^{\nu}{}_{\nu}- \Box\phi H^{\mu}{}_{\mu}+\partial_{\mu}\partial_{\nu}\phi H^{\mu\nu}\bigg),\label{stuckelberg graviton}
\end{equation}
where $F_{\mu\nu}$ is the usual field strength of $a_{\mu}$.

This result teaches us an important lesson. 
If we want to start from the massless fields $\phi$ and $a$ and couple them to symmetric tensor background fields, the corresponding currents can be read off \eqref{stuckelberg graviton}. Moreover, the starting point has to be different than simply the sum of the canonical kinetic terms for $\phi$ and $a$; total derivatives will play a role in this. Before presenting the details of this procedure, let us comment on the coupling of $H$ with $\phi$ and the apparent absence of the kinetic term for the scalar in \eqref{stuckelberg graviton}. As already explained in \cite{Arkani-Hamed:2002bjr}, this can be accounted for by a Weyl rescaling of the graviton, 
\begin{equation}
    H_{\mu\nu}\to \widehat{H}_{\mu\nu}= H_{\mu\nu}+\eta_{\mu\nu}\phi\,. \label{weyl 1}
\end{equation}
The effect of this is that the kinetic mixing of the scalar and the graviton disappears and the correct kinetic term and a mass term for the scalar field are generated. Then in the ``mass" $\to 0$ limit, the only term in the Stueckelberg action that contains the scalar field is  
\begin{equation}
    S_{\text{FP}}^{\text{St}}\supset \int\dd^{4}x\, \phi\,\Box \phi\,.
\end{equation}
Essentially, this is due to the fact that the mixing terms of $\phi$ and $H_{\mu\nu}$ in \eqref{stuckelberg graviton} generate the term $\phi R$, with $R$ the linearized curvature scalar for $H_{\mu\nu}$. The role of the scalar field here is very different from that of the usual Goldstone scalar. In the zero-coupling limit it decouples, unlike the zero coupling limit in gauge theory where the Goldstone boson kinetic term survives \cite{Arkani-Hamed:2002bjr}. We will call such fields ``secondary'', as further explained in Section \ref{sec:Discussion_Noether}.

\subsection{Tensor global symmetries \& gauging}

The Stueckelberg approach gives a quick way to identify the additional degrees of freedom that are needed to restore gauge invariance. However, it hides several important features, such as the structure of global symmetries, conserved currents, and anomalies. To reveal those features, we need a different starting point. Therefore, we start with the free fields $(\phi,a)$ and ask how we can couple a symmetric tensor field to the theory. 

Guided by the result of Section \ref{subsection massive graviton}, we consider the following, unconventional form of Maxwell theory: 
\begin{eqnarray}
    S[a,\phi] = \int\dd^{4}x\big(-\frac 12 \partial_{\mu}a_{\nu}\partial^{\mu}a^{\nu}+\frac 12 \partial^{\mu}a_{\mu}\partial^{\nu}a_{\nu} +\, \Box\phi\,\Box\phi-\partial_{\mu}\partial_{\nu}\phi\,\partial^{\mu}\partial^{\nu}\phi\big)\,. \label{symmetric Maxwell}
\end{eqnarray}
The kinetic term of $a_{\mu}$ is obtained by partial integration of the usual one and it is identical to it up to a total derivative{\footnote{This form of the Maxwell action was suggested in \cite{Boulanger:2015mka} in relation to exotic dual formulations of electromagnetism.}}. The scalar terms are all together just a total derivative, so they do not modify the classical field equations\footnote{Such four derivative terms appeared already in the effective field theory approach of \cite{Arkani-Hamed:2002bjr}, where it was also noticed that only with the choice of Fierz-Pauli coefficients they are not pathological.}. However, they are important for correctly identifying the conserved currents. Note that the $a_{\mu}$ terms are nothing more than $\dd\widetilde{a}\star\dd\widetilde{a}$ and the $\phi$ terms are $\dd\widetilde{\dd}\phi\star \dd\widetilde{\dd}\phi$, both encountered in the compact formulation of the Stueckelberg action. 

In this formulation, the Noether current corresponding to the global shift symmetry of $a_\mu$ acquires a different form than the ones of Section \ref{section forms}:
\begin{equation}
J^{(a)}_{\mu\nu}=\frac 12 (\partial_{\mu}a_{\nu}+\partial_{\nu}a_{\mu})-\eta_{\mu\nu}\partial^{\rho}a_{\rho} + \frac{1}{2} F_{\mu\nu}\,. 
\end{equation}
It is conserved on-shell,
\begin{equation}
\partial^{\mu}J^{(a)}_{\mu\nu}=0\,,
\end{equation}
due to Maxwell equations.
This current can be further decomposed into two independently conserved currents,
\begin{align}
J^{(a)}_{(\mu\nu)}&= \frac12 (\partial_{\mu}a_{\nu}+\partial_{\nu}a_{\mu} )-\eta_{\mu\nu}\partial^{\rho}a_{\rho},\\[4pt] 
   J^{(a)}_{[\mu\nu]}&= \frac12 F_{\mu\nu}\,, 
\end{align}
where $J^{(a)}_{(\mu\nu)}$ and $J^{(a)}_{[\mu\nu]}$ are respectively symmetric and antisymmetric.
While $J^{(a)}_{[\mu\nu]}$ is gauge invariant under a gauge transformation of the form $\delta a_{\mu}=-\partial_{\mu}\varphi$, the transformation of $J^{(a)}_{[\mu\nu]}$ does not vanish. We define the auxiliary current
\begin{equation}\label{eq:current_phi_aux}
J^{(\phi)}_{\mu\nu}= \partial_{\mu}\partial_{\nu}\phi- \eta_{\mu\nu}\Box\phi\,.
\end{equation}
This is also symmetric and conserved, since 
\begin{equation}
\partial^{\mu}J^{(\phi)}_{\mu\nu}=0\,,
\end{equation}
even off-shell. Its transformation under spacetime-dependent shifts $\delta\phi=\varphi$ is opposite to the one of the previous current. This motivates us to define the ``electric'' conserved current 
\begin{equation}
J^{\text{e}}_{\mu\nu}:=J^{(a)}_{(\mu\nu)}+J^{(\phi)}_{\mu\nu}\,,
\end{equation}
which is also invariant under gauge transformations of $a_{\mu}$.
The existence of this current\footnote{It is important to clarify that in the gauging procedure the coupling of the antisymmetric current $J^{(a)}_{[\mu\nu]}$ to a symmetric background field identically vanishes. This current is relevant in the discussion of Section \ref{section forms} where the current couples to a $2$-form background field.} signals the presence of a \emph{tensor global symmetry} with conserved current of type $(1,1)$, which we will denote as $U(1)_{e}^{(1,1)}$. 
 This is the current that we will couple to the symmetric 2-tensor background field $H_{\mu\nu}$. The minimal coupling is 
 \begin{eqnarray}
     J^{\text{e}}_{\mu\nu} H^{\mu\nu} = \partial_{\mu}a_{\nu}H^{\mu\nu}-\partial^{\rho}a_{\rho}H^{\mu}{}_{\mu}  +\,  \partial_{\mu}\partial_{\nu}\phi H^{\mu\nu}- \Box\phi H^{\mu}{}_{\mu}\,. \label{minimal coupling to graviton}
 \end{eqnarray}
 These couplings are precisely the ones we found in the Stueckelberg approach, Eq. \eqref{stuckelberg graviton}, which justifies that this is the right current. 
 
 Since we have identified symmetric currents, it is worth noting that their traces are 
 \begin{equation}
     J^{(a)}_{\mu}{}^{\mu}=-3\,\partial^{\mu}a_{\mu} \quad \text{and}\quad J^{(\phi)}_{\mu}{}^{\mu}=-3\,\Box \phi\,,
 \end{equation}
 which appear in the minimal coupling. This also shows that the status of $a$ and $\phi$ is different, as already seen through Weyl rescaling \eqref{weyl 1}. The trace of $J^{(a)}$ is the divergence of $a_{\mu}$ that appears in the Lorenz gauge condition, whereas the trace of $J^{(\phi)}$ is proportional to the d'Alembertian of $\phi$ that appears in the wave equation.   

 Gauging is performed in the usual way. Adding the minimal coupling \eqref{minimal coupling to graviton} to the original Lagrangian \eqref{symmetric Maxwell}, a ``seagull'' term quadratic in $H_{\mu\nu}$ makes the kinetic term invariant under the shift symmetry with parameter $\alpha_{\mu}$. Then we may add the kinetic term for $H_{\mu\nu}$, turning the background field into a dynamical one. It is worth recalling that in the massless limit, the theory at this level does not result in the linearized limit of general relativity (vDVZ discontinuity) \cite{vanDam:1970vg,Zakharov:1970cc}, unless higher-order terms in $H_{\mu\nu}$ are included \cite{Vainshtein:1972sx}. 

\subsection{Alternative view of the Noether symmetry}\label{sec:Discussion_Noether}

There is an equivalent yet seemingly different view of the symmetry transformations \eqref{eq:trans_H}-\eqref{eq:trans_phi_a}. We can rewrite them as
\begin{align}\label{eq:trans_Haphi}
    \delta H_{\mu\nu}&=\partial_{\mu}\alpha_{\nu}+\partial_{\nu}\alpha_{\mu}\,,\nonumber\\[4pt]
    \delta a_{\mu}&=\alpha_{\mu} - \partial_\mu{\varphi}\,,\\[4pt]
    \delta\phi&=\varphi\,.  \nonumber
\end{align}
This form is more natural from the Noether perspective. Firstly, we do not encounter a two-derivative transformation \eqref{eq:trans_H} and secondly the gauge transformation $\delta a_\mu = - \partial_\mu{\varphi}$ is manifest and hence the gauge invariance of the current seems more natural. Moreover, looking at the symmetries in this manner, the auxiliary symmetric current \eqref{eq:current_phi_aux} is nothing but the freedom in the definition of the Noether current which is unique up to an exact term \cite{Barnich:2001jy}. It means one can improve the symmetric conserved Noether current $J^{(a)}_{(\mu\nu)}$ by a divergence term, i.e.,
\begin{equation} 
J^{\text{e}}_{\mu\nu} = J^{(a)}_{(\mu\nu)} + \partial^\rho  k^{(\phi)}_{\rho\mu\nu}\,, 
\end{equation} 
where $k^{\rho\mu\nu}= k^{\rho(\mu\nu)}$ and $k^{(\rho\mu\nu)}=0$, subject to the constraint that $\partial^\mu J^{\text{e}}_{\mu\nu}=0$. From Eq. \eqref{eq:current_phi_aux}, we can see that
\begin{equation}
    k^{\rho\mu\nu} = \eta^{\rho(\mu} \partial^{\nu)}\phi -\eta^{\mu\nu} \partial^{\rho}\phi\,, 
\end{equation}
and it satisfies $\partial_{\mu}\partial_{\rho} k^{\rho\mu\nu}= \partial_{\nu}\partial_{\rho} k^{\rho\mu\nu}=0$ which guarantees that $\partial^\mu J^{\text{e}}_{\mu\nu}=0$. A similar discussion applies in the following sections, which we will not repeat.

There is an important distinction between the Stueckelberg field $a_\mu$ and the ``secondary" Stueckelberg field $\phi$, that can be understood from their transformation under the shift symmetry \eqref{eq:trans_Haphi}. The ``secondary" Stueckelberg field $\phi$ cannot become the Goldstone boson. This is due to the fact that its global shift symmetry has already been gauged by $a_\mu$. In fact, this is a general behavior; only those Stueckelberg fields can become Goldstone bosons as a result of symmetry breaking if in the corresponding set of symmetry transformations as in \eqref{eq:trans_Haphi} their shift symmetry is gauged by the background field and not by another Stueckelberg field. That is to say, the terminology ``secondary" Stueckelberg field indicates that it cannot become a Goldstone boson. Another way to understand this point is by noticing that after performing the Weyl transformation \eqref{weyl 1}, there will appear a mass term for the ``secondary" Stueckelberg field contrary to the other Stueckelberg field that is massless. It is well understood that only massless fields can become Goldstone bosons \cite{Weinberg_1996}.   
\section{Tensor shift symmetries II}
\label{section tensor GS 2}

\subsection{Stueckelberg action for the massive Curtright field}
\label{section massive curtright}

Climbing one step higher in the ladder of massive tensor fields, we encounter the case of a $(2,1)$ field. This was first studied as a generalized gauge field in \cite{Curtright:1980yk} and is often called the Curtright field. In components, the field is $T_{\mu\nu|\rho}$, the vertical line separating antisymmetric indices from the rest, namely the tensor field satisfies
\begin{equation}
    T_{(\mu\nu)|\rho}=0 \quad \text{and} \quad  T_{[\mu\nu|\rho]}=0\,.
\end{equation}
The massive theory for this field was also described in \cite{Curtright:1980yk}. As with the graviton, the coefficients of the different terms in the mass term must be carefully chosen to avoid propagating ghost degrees of freedom. Note that the massless Curtright field in 4D does not have any physical polarizations, whereas the massive one has 5 physical polarizations. Although our working dimension is $D=4$, the discussion in the following applies to any dimension.

The Lagrangian for the massive Curtright field is 
\begin{eqnarray}
  &&  S_{\text{Curt}}=-\frac 14\int\dd^4x \bigg(\partial_{\mu}T_{\nu\rho|\sigma}\partial^{\mu}T^{\nu\rho|\sigma}+2\partial^{\nu}T_{\nu\rho|}{}^{\rho}\partial^{\mu}T_{\mu\sigma|}{}^{\sigma}  -2 \partial_{\mu}T_{\nu\rho|}{}^{\rho}\partial^{\mu}T^{\nu\sigma|}{}_{\sigma} \nonumber \\[4pt] 
    && \hspace{100pt} -2 \partial^{\nu}T_{\nu\rho|\sigma}\partial_{\mu}T^{\mu\rho|\sigma} - \partial_{\mu}T_{\nu\rho|}{}^{\mu}\partial_{\sigma}T^{\nu\rho|\sigma} +4 T_{\nu\rho|}{}^{\rho}\partial_{\mu}\partial_{\sigma}T^{\mu\nu|\sigma}\nonumber\\[4pt] && \hspace{100pt} - T_{\mu\nu|\rho}T^{\mu\nu|\rho}+2T_{\mu\nu|}{}^{\nu}T^{\mu\rho|}{}_{\rho}\bigg) \nonumber\\[4pt]
    &&\hspace{28pt} =-\frac 12 \int\dd^{4}x\int_{\text{B}}\bigg(\dd T\star \dd T-T\star T\bigg)\,. 
    \label{massive Curtright}
\end{eqnarray} 
The complicated expression in components takes the exact same form in the compact formulation as for any other field, which significantly simplifies the analysis. Due to the mass term, the massive Curtright action is not invariant under the gauge symmetries 
\begin{equation} 
\delta T_{\mu\nu|\rho}=2\partial_{[\mu}s_{\nu]\rho}+2\partial_{[\mu}\beta_{\nu]\rho}-2\partial_{\rho}\beta_{\mu\nu}\,, 
\end{equation}
for a symmetric parameter $s_{\mu\nu}=s_{(\mu\nu)}$ and an antisymmetric parameter $\beta_{\mu\nu}=\beta_{[\mu\nu]}$\,. These are the gauge symmetries of the massless theory. As in the graviton case, it is useful to also consider the two-derivative transformations 
\begin{equation}
    \delta T_{\mu\nu|\rho}= 2\partial_{\rho}\partial_{[\mu}\alpha_{\nu]}\,,
\end{equation}
with a vector parameter $\alpha_{\mu}$, which are also not symmetries of the massive theory but they trivially are in the massless case. 

To restore gauge invariance under these symmetries, we must apply the Stueckelberg mechanism. In the present case, we need to introduce three massless fields: a graviton $h_{\mu\nu}$, a Kalb-Ramond field $b_{\mu\nu}$ and a 1-form $a_{\mu}$. This is completely fixed by the degree of freedom count of Section \ref{section dof} and there is no ambiguity. The shift symmetries for this host of Stueckelberg fields are 
\begin{equation}
    \delta h_{\mu\nu}=s_{\mu\nu}\,,\quad \delta b_{\mu\nu}=\beta_{\mu\nu}\,.\quad \delta a_{\mu}=\alpha_{\mu}\,.
\end{equation}
We will describe their additional gauge symmetries later. First, we redefine the Curtright field as 
\begin{equation}
    T_{\mu\nu|\rho} \to \mathring{F}_{\mu\nu|\rho}:=T_{\mu\nu|\rho}- 2\partial_{[\mu}h_{\nu]\rho}-2\partial_{[\mu}b_{\nu]\rho}+2\partial_{\rho}b_{\mu\nu}-2\partial_{\rho}\partial_{[\mu}a_{\nu]}\,.
\end{equation}
The redefined tensor $\mathring{F}_{\mu\nu|\rho}$ is invariant under all symmetries with spacetime dependent parameters $s_{\mu\nu}, \beta_{\mu\nu}$ and $\alpha_{\mu}$: 
\begin{equation}
    \delta \mathring{F}_{\mu\nu|\rho}=0\,.
\end{equation}
To simplify our analysis, we write the $(2,1)$ tensor $\mathring{F}_{\mu\nu|\rho}$ in the following compact form,
\begin{equation}
    \mathring{F}=T-\dd h-\dd\widetilde{\sigma}b +2 \widetilde{\dd}b-2\dd\widetilde{\dd}a\,,
    \label{ring F Curtright}
\end{equation}
where the operator $\widetilde{\sigma}$ is defined as 
\begin{equation}
    \widetilde{\sigma}=-\widetilde{\theta}^{\mu}\frac{\partial}{\partial \theta^{\mu}}\,,
\end{equation}
as a result of which $\widetilde{\sigma}b=b_{\mu\nu}\theta^{\mu}\widetilde{\theta}^{\nu}$ is a $(1,1)$ tensor.
Observing that $\dd \mathring{F}=\dd T-2\dd\widetilde{\dd}b$, we can write the massive Curtright action for the invariant redefined field: 
\begin{align}
   S_{\text{Curt}}^{\text{St}} = &-\frac 12 \int\dd^{4}x\int_{\text{B}}(\dd T-2\dd\widetilde{\dd}b)\star (\dd T-2\dd\widetilde{\dd}b) \\
   &+\frac 12 \int\dd^{4}x\int_{\text{B}} (T-\dd h -\dd\widetilde\sigma b+ 2\widetilde{\dd}b-2\dd\widetilde{\dd}a)\star (T-\dd h -\dd\widetilde\sigma b+ 2\widetilde{\dd}b-2\dd\widetilde{\dd}a)\,\nonumber.
\end{align}
Many terms are again total derivatives; these are $\dd T\star \dd\widetilde{\dd}b, \dd\widetilde{\dd}b\star \dd \widetilde{\dd}b, \dd\widetilde{\dd}a\star\dd h, \dd\widetilde{\dd}a\star\widetilde{\dd}b, \dd\widetilde{\dd}a\star \dd\widetilde{\dd}a, \dd\widetilde\sigma b\star\dd\widetilde\dd a, \dd\widetilde\sigma b\star \widetilde\dd b, \dd h\star\dd\widetilde{\sigma}b$ and $\dd h\star \widetilde{\dd}b$. After we drop the total derivatives, we are left with 
\begin{eqnarray}
   S_{\text{Curt}}^{\text{St}}&=& -\frac 12 \int\dd^{4}x\int_{\text{B}}\big(\dd T\star \dd T-T\star T\big) \nonumber \\[4pt] 
   &&+\, \frac 12\int\dd^{4}x\int_{\text{B}}\big(\dd h\star\dd h+4\widetilde{\dd}b\star\widetilde{\dd}b +\dd\widetilde\sigma b\star \dd\widetilde\sigma b\big) \nonumber \\[4pt] &&-\,\int\dd^{4}x\int_{\text{B}}\big(T\star \dd h-2T\star \widetilde{\dd}b+T\star \dd\widetilde\sigma b+2T\star \dd\widetilde{\dd}a\big)\,.
\end{eqnarray}
This is the final form of the Stueckelberg action for the massive Curtright field. To facilitate the analysis below, we write the coupling terms for the Stueckelberg fields in component form: 
\begin{align}
   S_{\text{Curt}}^{\text{St}}-S_{\text{Curt}}=  &\,  2\int \dd^{4}x  \bigg(\frac 14 \partial_{\mu}h_{\nu\rho}\partial^{\mu}h^{\nu\rho}-\frac 12 \partial_{\mu}h_{\nu\rho}\partial^{\nu}h^{\mu\rho} +\frac 12 \partial_{\mu}h^{\mu\nu}\partial_{\nu}h^{\rho}{}_{\rho} -\frac 14 \partial_{\mu}h^{\nu}{}_{\nu}\partial^{\mu}h^{\rho}{}_{\rho} \nonumber \\[4pt] &  +\frac 1{12} H_{\mu\nu\rho}H^{\mu\nu\rho} - \frac 12 T_{\mu\nu|}{}^{\mu}(\partial^{\nu}h_{\rho}{}^{\rho}-\partial^{\rho}h_{\rho}{}^{\nu})-\frac 12 T_{\mu\nu|\rho}\partial^{\mu}h^{\nu\rho} \nonumber\\[4pt] 
    &   -\frac 12 T_{\mu\nu|}{}^{\mu}\partial_{\rho}b^{\rho\nu} +\frac 12  T_{\mu\nu|\rho}(\partial^{\rho}b^{\mu\nu}+\partial^{\mu}b^{\nu\rho}) \nonumber \\[4pt] & - T_{\mu\nu|}{}^{\mu}(\partial^{\nu}\partial^{\rho}a_{\rho}-\Box a^{\nu})- T_{\mu\nu|\rho}\partial^{\rho}\partial^{\mu}a^{\nu}\bigg)\,, 
\end{align}
where the field strength of the Kalb-Ramond field is defined as 
\begin{equation}  H_{\mu\nu\rho}=\partial_{\mu}b_{\nu\rho}+\partial_{\rho}b_{\mu\nu}+\partial_{\nu}b_{\rho\mu}\,,
\end{equation}
and we observe that the kinetic term of the Kalb-Ramond field is correctly normalized relatively to the kinetic term for the graviton, with the characteristic $1/12$ coefficient. 

We observe that the linearized Einstein-Hilbert action is obtained as the lowest order term of a ``sigma model'' for the graviton. However, it does not come alone. It is accompanied by the Kalb-Ramond kinetic term and also with the auxiliary 1-form whose kinetic term does not appear in the above action. In this higher Stueckelberg mechanism, the 1-form plays the same secondary role as the scalar in Section \ref{subsection massive graviton}. It initially features a potentially pathological four-derivative kinetic term, which, however, turns out to be a total derivative. Its canonical kinetic term can be generated by an analog of the Weyl rescaling, where we redefine 
\begin{equation}
    T_{\mu\nu|\rho}\to \hat{T}_{\mu\nu|\rho}=T_{\mu\nu|\rho}+4\eta_{\rho[\mu}a_{\nu]}\,.
\end{equation}
The effect of this redefinition is that the kinetic mixing between $T_{\mu\nu|\rho}$ and $a_{\mu}$ disappears and instead a kinetic term appears for the 1-form $a_{\mu}$.
These considerations find a natural explanation in terms of gauging tensor global symmetries, as we show immediately below.

\subsection{Graviton \& Co. as Nambu-Goldstone bosons} \label{section graviton NG}

Building on the results of the previous section, we would like to see whether we can think of the graviton $h_{\mu\nu}$ as a Nambu-Goldstone boson for a spontaneously broken tensor global symmetry. One thing we know for sure is that for this to happen in a consistent way, we need to accompany it by the Kalb-Ramond field and the secondary 1-form field. However, we also know that we should be careful about the starting point, in particular about the action we should consider for these massless fields. 
Guided by the Stueckelberg approach, we start with the following action: 
\begin{align}
   & S[h,b,a]= \,S_{\text{LEH}}-\frac 14\int\dd^{4}x(\partial_{\rho}b_{\mu\nu}\partial^{\rho}b^{\mu\nu}-2\partial^{\mu}b_{\mu\nu}\partial_{\rho}b^{\rho\nu}) \nonumber\\[4pt] 
     & \qquad + \frac 18\int\dd^{4}x
    \bigg((\Box a_{\nu}-\partial_{\nu}\partial_{\mu}a^{\mu})(\Box a^{\nu}-\partial^{\nu}\partial^{\rho}a_{\rho})  -\,\partial_{\mu}\partial_{\nu}a_{\rho}\partial^{\nu}(\partial^{\rho}a^{\mu}-\partial^{\mu}a^{\rho})\bigg)\,. \label{LEH + KR}
\end{align}
The first term is just the usual linearized Einstein-Hilbert action, given by \eqref{fierz pauli} without the mass terms. The second term is an unorthodox way to rewrite the action for a Kalb-Ramond field, essentially after performing an integration by parts and dropping the boundary term. The sector that contains the 1-form $a_{\mu}$ is also unorthodox, it is just a complicated way to write a total derivative, and it does not contribute to the field equations. 

We would now like to identify a conserved current. First we can directly write the following Noether currents for each field:
\begin{eqnarray}
\label{J h}    J_{\mu\nu|\rho}^{(h)}&=&  \partial_{[\mu}h_{\nu]\rho}+\eta_{\rho[\mu}\partial_{\nu]}h_{\sigma}{}^{\sigma}-\eta_{\rho[\mu}\partial^{\sigma}h_{\nu]\sigma}\,,\\[4pt] 
   \label{J b}     J_{\mu\nu\rho}^{(b)}&=& \partial_{\rho}b_{\mu\nu}+\partial_{[\mu}b_{\nu]\rho}+\eta_{\rho[\mu}\partial^{\sigma}b_{\nu]\sigma}\,, \\[4pt] 
            J_{\mu\nu|\rho}^{(a)}&=& \partial_{\rho}\partial_{[\mu}a_{\nu]}+\eta_{\rho[\mu}\partial_{\nu]}\partial^{\sigma}a_{\sigma}-\eta_{\rho[\mu}\Box a_{\nu]}\,.
\end{eqnarray}
$J^{(h)}$ and $J^{(a)}$ are genuine $(2,1)$ tensors, since they are antisymmetric in the first two indices by construction and their fully antisymmetric part vanishes. $J^{(b)}$ on the other hand, has a fully antisymmetric part (note the absence of vertical index separator in \eqref{J b}), which would anyway vanish when contracted with an irreducible $(2,1)$ tensor. In fact, a genuine $(2,1)$ current for the Kalb-Ramond field may be defined by noting that $J^{(b)}_{[\mu\nu\rho]}=\frac 23 H_{\mu\nu\rho}$. Then we can subtract the fully antisymmetric part from \eqref{J b} and construct a genuine $(2,1)$ current that reads:
\begin{equation}
J_{\mu\nu|\rho}^{(b)}=\frac 13 \big(\partial_{\rho}b_{\mu\nu}-\partial_{[\mu}b_{\nu]\rho}+3\eta_{\rho[\mu}\partial^{\sigma}b_{\nu]\sigma}\big)\,.
\end{equation}
Henceforth, we will work with this conserved current.
Moreover, all $(2,1)$ currents defined above satisfy the following divergence conditions 
\begin{equation}
\partial^{\mu}J_{\mu\nu|\rho}^{(h,b,a)}=0=\partial^{\rho}J_{\mu\nu|\rho}^{(h,b,a)}\,.
\end{equation}
This means that they are all conserved currents in both sets of indices. Three of these conditions are satisfied on-shell, particularly 
\begin{eqnarray}
\partial^{\mu}J^{(h)}_{\mu\nu|\rho}&=&\frac 12(\Box h_{\nu\rho}-\partial^{\mu}\partial_{\nu}h_{\mu\rho}+\eta_{\rho\mu}\partial^{\mu}\partial_{\nu}h_{\sigma}{}^{\sigma}\nonumber\\[4pt] && -\,\eta_{\nu\rho}\Box h_{\mu}{}^{\mu} -\partial_{\rho}\partial^{\sigma}h_{\nu\sigma}+\eta_{\nu\rho}\partial^{\mu}\partial^{\sigma}h_{\mu\sigma})\,,
\end{eqnarray}
the right hand side being the linearized form of Einstein equations, and similarly, 
\begin{equation}
    \partial^{\rho}J^{(b)}_{\mu\nu|\rho}=\partial^{\rho}J^{(b)}_{\rho\mu|\nu}= \frac 13 \partial^{\rho}H_{\mu\nu\rho}\,,
\end{equation}
the right hand side being the field equation for the Kalb-Ramond field. The rest of the conservation laws are identically satisfied off-shell, essentially following from the various Bianchi identities. 

To compile the above into a complete current, we look at their behavior under gauge symmetries. This time we look into the cascade of gauge transformations given by 
\begin{eqnarray}
 \label{gauge a}   \delta a_{\mu}&=&-\alpha_{\mu}+\partial_{\mu}\varepsilon\,, \\[4pt] 
 \label{gauge b}   \delta b_{\mu\nu}&=& 2\partial_{[\mu}\alpha_{\nu]}\,, \\[4pt]
 \label{gauge h}   \delta h_{\mu\nu}&=& 2\partial_{(\mu}\alpha_{\nu)} +  \partial_{\mu}\partial_{\nu}\varepsilon\,.
\end{eqnarray}
We find that under these transformations 
\begin{equation} 
\delta J^{(h)}= \delta J^{(b)}= -\delta J^{(a)}=\partial_{\rho}\partial_{[\mu}\alpha_{\nu]}+\eta_{\rho[\mu}\partial_{\nu]}\partial^{\sigma}\alpha_{\sigma}-\eta_{\rho[\mu}\Box\alpha_{\nu]}\,.
\end{equation} 
We can then define the doubly conserved and gauge invariant $(2,1)$ current
\begin{equation}
    J_{\mu\nu|\rho}:=J_{\mu\nu|\rho}^{(h)}+  J_{\mu\nu|\rho}^{(b)}-2 J_{\mu\nu|\rho}^{(a)}\,. \label{current graviton + KR} 
\end{equation}
This results in a conserved current of type $(2,1)$ that corresponds to a tensor global symmetry which we denote as $U(1)^{(2,1)}$. We can couple this current to a background field, which is now a Curtright field $T_{\mu\nu|\rho}$ as 
\begin{equation}
    J_{\mu\nu|\rho}T^{\mu\nu|\rho}\,.  \label{minimal h,b T}
\end{equation}
This results in the desired couplings. Gauging requires the introduction of a seagull term for the Curtright field and a kinetic sector for it, precisely corresponding to the actions we described in the previous section. We will discuss a magnetic current and a 't Hooft anomaly in Section \ref{section anomalies}.

Since the currents are tensors of mixed symmetry, we can also compute their traces. They are 
\begin{eqnarray}
  J^{(h)}_{\mu\nu}{}^{\mu}&=&\partial_{\nu}h_{\mu}{}^{\mu}-\partial^{\mu}h_{\mu\nu}\,, \\[4pt] J^{(b)}_{\mu\nu}{}^{\mu}&=&-\,2\partial^{\mu}b_{\mu\nu}\,, \\[4pt]  J^{(a)}_{\mu\nu}{}^{\mu}&=&-\,\partial^{\mu}F_{\mu\nu}\,.  
\end{eqnarray}
Once again, for $h_{\mu\nu}$ and $b_{\mu\nu}$ the traces contain the terms that appear in gauge conditions (transverse traceless and Lorenz, respectively) and for the secondary field $a_{\mu}$, whose kinetic coupling to the background field may be eliminated by a rescaling, the trace is proportional to its equation of motion.  

It is instructive to compare our results to previous approaches on the graviton as a Nambu-Goldstone boson. The conserved and gauge-invariant current \eqref{current graviton + KR}, when the Kalb-Ramond and the vector fields are set to zero, reduces to the current \eqref{J h}, which was calculated in \cite{Hull:2024bcl} as
\begin{equation}\label{eq:Hulletal_current}
J^{({h})}_{\mu \nu |}{}^{\rho} = 3 \delta_{\mu \nu \sigma}^{\kappa\lambda \rho} \partial_{[\kappa}h_{\lambda]}^{\sigma}\,,
\end{equation}
where $\delta_{\mu \nu \sigma}^{\kappa\lambda \rho}$ is the generalized Kronecker delta. However, from the standpoint of our approach, this does not constitute a consistent truncation, as the matching of physical polarizations in the Stueckelberg mechanism would fail. Furthermore, the current \eqref{eq:Hulletal_current} is not gauge-invariant under the shift symmetry of $h_{\mu\nu}$, i.e., \eqref{gauge h} with $\varepsilon= 0$.

Finally, as shown in \cite{Hull:2024bcl}, the current $J^{(h)}$ reduces further to the current found in \cite{Hinterbichler:2022agn}, which corresponds to a special case of shift symmetry where the symmetric shift parameter is written as a divergence of a $(2,1)$ parameter,
\begin{equation}
\delta h_{\mu \nu} = 2 \partial^{\rho} \Lambda_{\rho(\mu|\nu)}\,,
\end{equation}
where $\Lambda_{\rho \mu |\nu}$ is a $(2,1)$ parameter. Here, some spatial dependence of the shift is allowed and the corresponding current is identified with the linearized Riemann tensor. Our formulation thus extends the approaches of \cite{Hull:2024bcl} and \cite{Hinterbichler:2022agn}, highlighting the importance of the accompanying fields in maintaining consistency of the number of degrees of freedom and of the gauge symmetries.

\section{Exotic Stueckelberg mechanism}
\label{section exotic}

The final example we would like to work out is an exotic version of Stueckelberg mechanism, in the sense that the set of Stueckelberg bosons contains an exotic massless field of type $(2,1)$. It also contains a graviton as ``secondary" Stueckelberg field, which as previously discussed cannot become a Goldstone boson in the present case. Since the basic steps of the procedure are the same as in previous cases, we will be brief. The starting point in this case is a massive field with the same symmetries as the Riemann tensor, a $(2,2)$ tensor field $\Omega$. We cannot work in 4D any longer, since a $(2,2)$ does not have any physical degrees of freedom in 4D, regardless if it is massive or massless. However, the massive field has 10 polarizations in 5D and the action that does not propagate any ghosts is
\begin{eqnarray}
 S_{(2,2)}=   -\frac 12 \int\dd^{5}x\int_{\text{B}}\bigg(\dd\Omega\star \dd \Omega-\Omega\star\Omega \bigg)\,,
\end{eqnarray}
exactly of the same form as in all previous cases. The mass term in the present case takes the following explicit form in components: 
\begin{eqnarray}
\int_{\text{B}}\Omega\star\Omega=-\frac 14 \big(\Omega_{\mu\nu}{}^{\mu\nu}\Omega_{\kappa\lambda}{}^{\kappa\lambda}+\Omega_{\mu\nu}{}^{\kappa\lambda}\Omega^{\mu\nu}{}_{\kappa\lambda}-4\Omega_{\mu\nu}{}^{\mu\kappa}\Omega^{\lambda\nu}{}_{\lambda\kappa}\big)\,.\qquad 
\end{eqnarray}
The gauge symmetry of the massless theory, which does not survive in the massive theory, is 
\begin{equation}
\delta \Omega_{\mu\nu|\kappa\lambda} = \partial_{[\mu}\varepsilon_{\kappa\lambda|\nu]}+  \partial_{[\kappa}\varepsilon_{\mu\nu|\lambda]}+\partial_{[\mu}\partial_{[\kappa}s_{\nu]\lambda]}
\end{equation}
where aside the $(2,1)$ gauge parameter $\varepsilon_{\mu\nu|\kappa}$ we included the unconventional, cohomological part with a $(1,1)$ parameter $s_{\mu\nu}$. To restore the symmetry, we introduce a massless Curtright field $t$ and a massless graviton $h$. Then the tensor 
\begin{equation}
    \mathring{F}= \Omega-\dd\widetilde{t}-\widetilde{\dd}t-\dd\widetilde{\dd}h
\end{equation}
is gauge invariant and the Stueckelberg action is 
\begin{eqnarray}
    S_{(2,2)}^{\text{St}}&=&-\frac 12 \int\dd^{5}x\int_{\text{B}}\bigg(\dd\Omega\star\dd\Omega-2\dd\widetilde{t}\star\dd\widetilde{t}-\Omega\star\Omega\bigg) \nonumber\\[4pt] 
   && -\,\int\dd^{5}x\int_{\text{B}}\bigg(2 \Omega\star\widetilde{\dd}t+\Omega\star \dd\widetilde{\dd}h\bigg)\,.
   \label{2,2 stuck}
\end{eqnarray}
The component form of the couplings of $\Omega$ with the massless fields is   
\begin{align}
 \int_{\text{B}} 2 \Omega\star\widetilde{\dd}t= &~\frac 14(\Omega_{\mu\nu}{}^{\mu\nu}\partial_{\kappa}t^{\kappa\lambda}{}_{\lambda}-2\Omega_{\mu\nu}{}^{\mu\kappa}\partial_{\kappa}t^{\nu\lambda}{}_{\lambda} + 2\Omega_{\mu\nu}{}^{\mu\kappa}\partial_{\lambda}t^{\nu\lambda}{}_{\kappa}+\Omega_{\mu\nu}{}^{\kappa\lambda}\partial_{\kappa}t^{\mu\nu}{}_{\lambda})\,,\\[4pt] 
 \int_{\text{B}}\Omega\star \dd\widetilde{\dd}h= &\,\frac 14(\Omega_{\mu\nu}{}^{\mu\kappa}\partial^{\lambda}\partial^{\nu}h_{\kappa\lambda}-\Omega_{\mu\nu}{}^{\mu\kappa}\Box h_{\kappa}{}^{\nu}+\Omega_{\mu\nu}{}^{\mu\kappa}\partial_{\kappa}\partial_{\lambda}h^{\lambda\nu} - \Omega_{\mu\nu}{}^{\mu\kappa}\partial_{\kappa}\partial^{\nu}h^{\lambda}{}_{\lambda} \nonumber\\[4pt] &+\Omega_{\mu\nu}{}^{\kappa\lambda}\partial_{\kappa}\partial^{\mu}h_{\lambda}{}^{\nu} + \frac 12 \Omega_{\mu\nu}{}^{\mu\nu}\Box h_{\kappa}{}^{\kappa}-\frac 12 \Omega_{\mu\nu}{}^{\mu\nu}\partial_{\kappa}\partial_{\lambda}h^{\kappa\lambda})\,.
    \label{2,2 couplings}
\end{align}
The graviton now plays a different, third role in comparison to the previous sections. It is the analog of the scalar field $\phi$ in Section \ref{subsection massive graviton} and of the 1-form $a_{\mu}$ of Section \ref{section massive curtright}. It comes with a would-be pathological kinetic term which however is a total derivative and its canonical kinetic term appears only after a redefinition of the $(2,2)$ tensor field that eliminates its kinetic mixing terms with $h_{\mu\nu}$, of the form  
\begin{equation}
    \Omega_{\mu\nu|\kappa\lambda}\to \widehat{\Omega}_{\mu\nu|\kappa\lambda}=\Omega_{\mu\nu|\kappa\lambda}+2\eta_{[\mu[\kappa}h_{\nu]\lambda]}
\end{equation}
The reverse approach of gauging proceeds in the same way as before, starting from the suitable Lagrangian for the massless fields read off \eqref{2,2 stuck} and identifying the conserved and gauge invariant $(2,2)$ current $J_{\mu\nu|\kappa\lambda}$ read off the couplings \eqref{2,2 couplings}. 
This is given as 
\begin{equation}
J_{\mu\nu|\kappa\lambda}= J_{\mu\nu|\kappa\lambda}^{(t)}+J_{\mu\nu|\kappa\lambda}^{(h)}\,,
\end{equation}
where 
\begin{eqnarray}
J_{\mu\nu|\kappa\lambda}^{(t)}&=&\partial_{[\kappa}t_{\mu\nu|\lambda]}-2\eta_{[\mu[\lambda}\partial^{\rho}t_{\nu]\rho|\kappa]} + 2\eta_{[\mu[\lambda}\partial_{\kappa]}t_{\nu]\rho|}{}^{\rho}+\eta_{[\kappa[\mu}\eta_{\lambda]\nu]}\partial^{\rho}t_{\rho\sigma|}{}^{\sigma}\,, \\[4pt] 
J_{\mu\nu|\kappa\lambda}^{(h)}&=& \partial_{[\mu}\partial_{[\kappa}h_{\nu]\lambda]}+\eta_{[\mu[\lambda}\partial_{\nu]}\partial_{\kappa]}h_{\rho}{}^{\rho}+\eta_{[\mu[\lambda}\Box h_{\nu]\kappa]} - \eta_{[\mu[\lambda}\partial_{\kappa]}\partial^{\rho}h_{\nu]\rho}-  \eta_{[\mu[\lambda}\partial_{\nu]}\partial^{\rho}h_{\kappa]\rho}\nonumber \\ 
&& +\,\frac 12 \eta_{[\mu[\kappa}\eta_{\nu]\lambda]}(\Box h_{\rho}{}^{\rho}-\partial_{\rho}\partial_{\sigma}h^{\rho\sigma})\,, \label{J h 2,2}
\end{eqnarray}
and it is doubly conserved and gauge invariant. (Note that the antisymmetrization in both formulas is taken only in the pairs of indices $[\mu\nu]$ and $[\kappa\lambda]$, as should be clear from the structure of the currents.) The precise form of these currents would have been hard to identify without the power of the Stueckelberg mechanism, combined with the compact notation based on graded coordinates. Nevertheless, the current associated to the graviton is very suggestive of curvature tensors. Indeed we may simply write it in terms of the linearized Riemann, Ricci and curvature scalar tensors as 
\begin{align}
J^{(h)}_{\mu\nu|\kappa\lambda}&=-\frac 12 S_{\mu\nu|\kappa\lambda}\,,\quad \label{current h}
\end{align}
where we defined the following covariant 4-tensor with the symmetries of the Riemann tensor: 
\begin{equation}
\label{S tensor}
S_{\mu\nu|\kappa\lambda}=R_{\mu\nu|\kappa\lambda}+\frac 12 (\eta_{\mu\kappa}\eta_{\nu\lambda}-\eta_{\mu\lambda}\eta_{\nu\kappa})R +\eta_{\mu\lambda}R_{\nu\kappa}-\eta_{\mu\kappa}R_{\nu\lambda}+\eta_{\nu\kappa}R_{\mu\lambda}-\eta_{\nu\lambda}R_{\mu\kappa}\,,
\end{equation}
whose linearization appears in \eqref{J h 2,2}.
The current satisfies the covariant divergence condition 
 \begin{equation}
\nabla^{\mu}J^{(h)}_{\mu\nu|\kappa\lambda}=0\,,
 \end{equation}
 and by symmetry this holds for the divergence on the other set of indices. The vanishing of the divergence is easily proven using the contracted Bianchi identities 
 \begin{equation}
\nabla^{\lambda}R_{\mu\nu\kappa\lambda}=\nabla_{\nu}R_{\mu\kappa}-\nabla_{\mu}R_{\nu\kappa} \quad \text{and} \quad \nabla^{\mu}R_{\mu\nu}=\frac 12 \nabla_{\nu}R\,.
 \end{equation}
 This justifies in a direct way that it is a conserved current in the context of this section. Note that neither the Riemann tensor nor the Weyl tensor have vanishing covariant divergence and indeed the current is none of them \footnote{One may note that the form of the current is given by the formula for the Weyl tensor in 3D, which vanishes identically in 3D but not in higher dimensions. On the other hand, the Weyl tensor in 5D is not given by this expression. The tensor $S_{\mu\nu|\kappa\lambda}$ is mentioned as the starting point in defining Riemann-Lovelock invariants in \cite{Kastor:2012se}.}. Moreover, the trace of the current is 
 \begin{equation}
\eta^{\mu\kappa}J^{(h)}_{\mu\nu|\kappa\lambda}=G_{\nu\lambda}\,
 \end{equation}
 where $G_{\mu\nu}=R_{\mu\nu}-\frac 12  g_{\mu\nu}R$ is the Einstein tensor. In other words, the tensor given in \eqref{S tensor} is to the Einstein tensor what the Riemann tensor is to the Ricci tensor, and it is in addition covariantly conserved. This agrees with the results of the previous sections, where the trace of the current of the secondary field is a field equation.  

\section{'t Hooft anomaly for the graviton}
\label{section anomalies}

There is one thing that we have swept under the carpet in the previous three sections. In the case of differential forms, we discussed that there are two currents, one electric and one magnetic, and that they give rise to a 't Hooft anomaly. This raises the question: \emph{does a 't Hooft anomaly arise for tensor global symmetries?}

We will work with the most interesting case of the Nambu-Golstone graviton of Section \ref{section massive curtright}. First, we note that apart from minimal coupling of the conserved current, one may also consider nonminimal couplings \cite{Seiberg:2019vrp}. To see this, we note that beside the one-derivative current \eqref{J h} there exist the following electric and magnetic two-derivative currents, corresponding to the Riemann tensor and its \emph{ordinary} Hodge dual in the first set of indices:
\begin{eqnarray}
J^{\text{e}}_{\mu\nu|\rho\sigma}&=&\partial_{[\mu}\partial_{[\rho}h_{\nu]\sigma]}\,, \\ 
J^{\text{m}}_{\mu\nu|\rho\sigma}&=&\ast J^{\text{e}}_{\mu\nu|\rho\sigma}=\frac 12 \epsilon_{\mu\nu}{}^{\kappa\lambda}\partial_{k}\partial_{[\rho}h_{\sigma]\lambda}\,. 
\label{magnetic current}
\end{eqnarray}
This is expected in view of the electric/magnetic duality in linearized gravity described in \cite{Hull:2001iu}, which exchanges these two currents. The existence of this duality also points in the direction of a 't Hooft anomaly in the spirit of our analysis, since it should be impossible to gauge at the same time the theory in both duality frames. 

The above currents are conserved on-shell and gauge invariant. This may be easily seen through their geometric form, $J^{\text{e}}=\dd\widetilde{\dd}h$ and $J^{\text{m}}=\ast\dd\widetilde{\dd}h$. Gauge invariance simply follows from the fact that the de Rham operator is nilpotent. Conservation, on the other hand, requires us to compute the divergence in both slots of the mixed-symmetry tensor. This is implemented using the dual (codifferential) operators $\dd^{\dagger}=\ast\,\dd\,\ast$ and $\widetilde{\dd}^{\dagger}=\widetilde{\ast}\,\widetilde{\dd}\,\widetilde{\ast}$. Then we find the desired continuity equations for both currents
\begin{equation}
    \dd^{\dagger}J^{\text{e,m}}=0= \widetilde{\dd}^{\dagger}J^{\text{e,m}}\,.
\end{equation}
This follows from the differential Bianchi identity for the Riemann tensor in the case of the magnetic current, and it holds off-shell. For the electric current, conservation holds on-shell and it follows from the following identities \cite{Chatzistavrakidis:2019len}
\begin{equation}
\dd^{\dagger}=\widetilde{\dd}\,\text{tr}+\text{tr}\,\widetilde{\dd} \quad \text{and} \quad \widetilde{\dd}^{\dagger}=\dd\,\text{tr}+\text{tr}\,\dd\,,
\end{equation}
where $\text{tr}$ is the usual trace and $\text{tr}\,\dd\widetilde{\dd}h=0$ are the linearized Einstein equations.

To couple these currents to background fields $T^{\text{e}}$ and $T^{\text{m}}$ respectively, which are both $(2,1)$ tensor fields in this case, we need a $(2,2)$ object. This is formed naturally as $\dd\widetilde{T}$, which arises naturally from Eq. \eqref{ring F Curtright}, since 
\begin{equation}
\dd\widetilde{\mathring{F}}=\dd\widetilde{T}-\dd\widetilde{\dd}h-\dd\widetilde{\dd}\sigma\widetilde{b}\,.
\end{equation}
We note that in this formalism the kinetic term for the Curtright field can also be written as the Berezin integral of $\dd\widetilde{T}\star\dd\widetilde{T}$, which gives the same result, up to a factor, as the one of $\dd T\star\dd T$. Then the nonminimal couplings of $h_{\mu\nu}$ to $T^{\text{e}}_{\mu\nu|\rho}$ and $T^{\text{m}}_{\mu\nu|\rho}$ that should be added to the action \eqref{LEH + KR} besides the minimal couplings \eqref{minimal h,b T} are 
\begin{eqnarray}
L_{\text{e}}&=&\int_{\text{B}}\dd\widetilde{T}^{\text{e}}\star J^{\text{e}}=\int_{\text{B}}\dd\widetilde{T}^{\text{e}}\star \dd\widetilde{\dd}h\,,\\[4pt]
     L_{\text{m}}&=& \int_{\text{B}}\dd\widetilde{T}^{\text{m}}\star J^{\text{m}}=\int_{\text{B}}\dd\widetilde{T}^{\text{m}}\star\ast\,\dd\widetilde{\dd}h\,.
\end{eqnarray}
Performing the Berezin integration, we get the following component expressions for the electric and magnetic couplings,
\begin{eqnarray}
L_{\text{e}}&=&-\partial^{[\mu}T_{\text{e}}^{\kappa\lambda|\nu]}S_{\mu\nu|\kappa\lambda}
\\[4pt]
    L_{\text{m}}&=& -\,\epsilon^{\nu_1\nu_2\nu_3\nu_4}\partial^{\kappa}T^{\text{m}}_{\nu_1\nu_2}{}^{\lambda}\partial_{\nu_3}\partial_{[\kappa}h_{\lambda]\nu_4}\,,
\end{eqnarray}
where $S_{\mu\nu|\kappa\lambda}$ is given by the linearization of \eqref{S tensor}. Since its divergence vanishes, $L_{\text{e}}$ is a total derivative, which is not the case for $L_{\text{m}}$. 
Under the background gauge transformation of the magnetic background field $T^{\text{m}}$, the magnetic coupling transforms to a total derivative and drops out under spacetime integration. On the other hand, $h_{\mu\nu}$ is inert under this transformation, but transforms under the electric background gauge transformation with the shift $s_{\mu\nu}$. Then we see that under this electric transformation 
\begin{equation}
    \delta_{\text{e}}L_{\text{m}}=-\,\epsilon^{\nu_1\nu_2\nu_3\nu_4}\partial^{\kappa}T^{\text{m}}_{\nu_1\nu_2}{}^{\lambda}\partial_{\nu_3}\partial_{[\kappa}s_{\lambda]\nu_4}\,.
\end{equation}
Having identified all these, we can now improve the discussion of Section \ref{section graviton NG}. Starting from the action \eqref{LEH + KR}, we can couple minimally the conserved current \eqref{current graviton + KR} as in \eqref{minimal h,b T} with $T=T^{\text{e}}$, and add the required seagull term for the electric Curtright background field. In addition, we may add to this the new couplings $L_{\text{m}}$ to the magnetic background field. The complete action of the graviton, Kalb-Ramond field and vector field coupled to both background fields is 
\begin{equation}
    S[h, b, a, T_{\text{e}},T_{\text{m}}]=S[h, b, a] + \int\dd^{4}x \,\big(J_{\mu\nu|\rho}T_{\text{e}}^{\mu\nu|\rho}-\epsilon_{\kappa\lambda}{}^{\mu\nu}J_{\mu\nu|\rho\sigma}\partial^{\rho}T_{\text{m}}^{\kappa\lambda|\sigma}\big)+S_{\text{ct}}[T_{\text{e}}],
\end{equation}
where $S[h, b, a]$ is the action \eqref{LEH + KR} (linearized Einstein-Hilbert plus kinetic term of the Kalb-Ramond field), $J_{\mu\nu|\rho}$ is given by \eqref{current graviton + KR}, $J_{\mu\nu|\kappa\lambda}$ is the magnetic current \eqref{magnetic current}, and $S_{\text{ct}}[T_{\text{e}}]$ is the counterterm term for the electric background field, given by the third line of \eqref{massive Curtright}. 
This action is invariant under magnetic background gauge transformations, 
\begin{equation}
    \delta_{\text{m}}S=0\,,
\end{equation}
it is, however, not invariant under electric ones 
\begin{eqnarray}
    \delta_{\text{e}}S&=&\int\dd^4 x\, \epsilon_{\kappa\lambda}{}^{\mu\nu}\partial^{\rho}T_{\text{m}}^{\kappa\lambda|\sigma}\partial_{\mu}\partial_{[\sigma}s_{\rho]\nu} \nonumber\\[4pt] &=& \int\dd^4 x\, \epsilon_{\kappa\lambda}{}^{\mu\nu}s_{\nu[\rho}\partial_{\sigma]}\partial_{\mu}\partial^{\rho}T_{\text{m}}^{\kappa\lambda|\sigma}\,,
\end{eqnarray}
where in the second line we performed a partial integration to bring it to a form directly comparable to the analogous one for $p$-forms in Eq. \eqref{anomaly p form}. This constitutes the 't Hooft anomaly for the graviton and it would appear in the partition function. As long as $T^{\text{m}}$ is a source and not a dynamical field, we can circumvent this by turning this background field off. The anomaly tells us though that we cannot promote both background fields to dynamical fields simultaneously. 

\section{Conclusions \& Outlook}\label{section conclusions}
In this paper we studied tensor fields as Goldstone bosons of spontaneously broken mixed-symmetry tensor global symmetries. These are shift symmetries of tensor fields, with parameters that share the same Young tableaux symmetry as the tensor field itself. By exploring the relationship between the Stueckelberg mechanism and the realization of Nambu-Goldstone bosons in the context of higher-form and mixed-symmetry tensor fields, we developed a systematic method for identifying gauge-invariant and conserved Noether currents corresponding to global tensor shift symmetries. These Noether currents are constructed as first derivatives of the fields, aligning with the gauge theory perspective on Noether currents for $p$-form fields. By extending the traditional framework of $p$-form gauge symmetries to include $(p,q)$-form symmetries, we explained how gauge invariance is restored in theories involving massive tensor fields and how these fields manifest themselves as Goldstone bosons in the spontaneous breaking of global symmetries. 

We demonstrated the use of this approach in three contexts. For the massive Fierz-Pauli theory, we identified the Stueckelberg fields required to restore gauge invariance—a massless $1$-form and a scalar—thereby revealing novel conserved, gauge-invariant symmetric currents associated with higher tensor global symmetries. A central focus of our work was addressing the graviton as a Goldstone boson, bridging gaps in existing constructions where currents were either built as second derivatives of fields or lacked gauge invariance when constructed from first derivatives. To resolve this, we considered the massive Curtright theory, where the interplay between gauge invariance and symmetry breaking underscored the necessity of multiple Stueckelberg fields—namely, a graviton, a Kalb-Ramond field, and a vector field—to fully restore gauge invariance. In this framework, the graviton and Kalb-Ramond field naturally emerge as Goldstone bosons of spontaneously broken symmetric and antisymmetric shift symmetries. The inclusion of the Kalb-Ramond field proved essential for constructing a gauge-invariant current defined in terms of the first derivatives of fields. To further demonstrate the generality of our construction for arbitrary $(p,q)$-mixed tensor symmetries, we constructed the Noether currents for the mixed-tensor shift symmetries of Stueckelberg fields in the massive $(2,2)$-tensor field theory.

Notably, our approach assigned a triple role to a symmetric 2-tensor, one for each of the cases described above. It served as a massive tensor field (respectively, a background field to which symmetric conserved 2-tensor currents couple), as a Goldstone boson accompanied by a Kalb-Ramond field when the background field is a $(2,1)$ tensor, and as a secondary Stueckelberg field when the backround field is a $(2,2)$ tensor. Although we always referred to this field as ``the graviton'', one may think of it more broadly as a symmetric tensor gauge field, especially in view of the usefulness of such fields in the context of the physics of subdimensional particles \cite{Pretko:2020cko}.

Additionally, electric-magnetic duality in tensor gauge theories provides a robust framework for studying 't Hooft anomalies of tensor global symmetries. In this work, we used nonminimal couplings of the currents to background fields to describe a 't Hooft anomaly for the graviton within our framework. However, due to the complexity arising from nonminimal couplings, deriving an anomaly polynomial for this 't Hooft anomaly is challenging.

Several questions remain open for future investigation. The first pertains to the description of the 't Hooft anomaly via a minimal coupling of gauge-invariant currents to background fields. A key difficulty here lies in constructing a gauge-invariant and fully conserved magnetic current. In a broader context, theories with mixed tensor symmetries often exhibit exotic dualities \cite{Boulanger:2015mka,Boulanger:2024lwk,Chatzistavrakidis:2019bxo}, and understanding the 't Hooft anomaly in these scenarios remains an unresolved issue. A second question concerns the construction of gauge-invariant topological defects charged under mixed-symmetry tensor global symmetries. For $p$-form symmetries, one can construct Wilson loops and higher-dimensional defects by integrating gauge-invariant currents over a codimension $p+1$ surface. However, for $(p,q)$-mixed tensor symmetries, naively integrating gauge-invariant currents over a codimension $p+1$ surface does not yield a topological object. In the specific case of graviton shift symmetry, the current approach of reducing the non-gauge-invariant $(2,1)$ current to a $2$-form gauge-invariant current and then integrating it to construct topological defects is unsatisfactory. A more direct approach, building on the gauge-invariant currents developed in this work, would be of interest. In relation to this, it would be interesting to understand the bound on dimensions where spontaneous symmetry breaking is possible in these cases, essentially an analog of the Coleman-Mermin-Wagner theorem. Finally, going beyond the linear theory is another challenge. In this respect, it is worth mentioning the recent proposal of a generalized symmetry in nonlinear gravity \cite{Cheung:2024ypq, Gomez-Fayren:2024cpl} and the generalized Stueckelberg mechanism of \cite{Ho:2012nt} that applies to non-Abelian 2-forms.

\acknowledgments

The authors are grateful to Giorgos Karagiannis and Jan Rosseel for participation in earlier stages of this work and for enlightening discussions. Useful discussions with Oleg Antipin are gratefully acknowledged. This work is funded by the Croatian Science Foundation project ``HigSSinGG — Higher Structures and Symmetries in Gauge and Gravity Theories'' (IP-2024-05-7921) and by the European Union — NextGenerationEU. The work of Ath. Ch. is also supported by the Ulam Programme of the Polish National Agency for Academic Exchange.

\bibliography{Stueckelberg_v2.bib} 
\bibliographystyle{utphys}

\end{document}